\documentclass[pre,aps,showpacs,12pt,reprint,superscriptaddress]{revtex4-1}

\usepackage[T1]{fontenc}
\usepackage[utf8]{inputenc}
\usepackage[english]{babel}
\usepackage{graphicx, color}
\usepackage{amssymb, amsmath, amsfonts, amsthm,stmaryrd}
\usepackage{dsfont}
\usepackage{bm}
\usepackage{ifthen}
\usepackage{mathrsfs}
\usepackage{ulem}

\everymath{\displaystyle}

\usepackage{hyperref}

\newcommand{\f}[2]{\mathchoice%
			{\dfrac{#1}{#2}}
	    	{\dfrac{#1}{#2}}
			{\frac{#1}{#2}}
			{\frac{#1}{#2}}}

\newcommand{\ddf}[3][]{\ifthenelse{\equal{#1}{}}{\ensuremath{\f{\dd#2}{\dd#3}}}
{\ensuremath{\f{\dd^{#1}#2}{\dd{#3}^{#1}}}}}
\newcommand{\Dp}[3][]{\ifthenelse{\equal{#1}{}}{\ensuremath{\f{\partial#2}{\partial#3}}}
  {\ensuremath{\f{\partial^{#1}#2}{\partial{#3}^{#1}}}}}

\renewcommand{\hat}[1]{\ensuremath{\widehat{#1}}}
\renewcommand{\tilde}[1]{\ensuremath{\widetilde{#1}}}

\newcommand{\avg}[1]{\ensuremath{\left\langle #1 \right\rangle}}
\newcommand{\expec}[1]{\ensuremath{{\rm I\kern-.3em \rm E}\left[#1\right]}}
\newcommand{\prob}[1]{\ensuremath{\pi\left(#1\right)}}

\newcommand{\e}[1]{\ensuremath{{}_{\text{#1}}}}

\renewcommand{\rm}[1]{\mathrm{#1}}
\newcommand{\T}{\rm{T}}

\newcommand{\eg}{\textit{e}.\textit{g}. }

\newcommand{\dd}{\mathrm{d}} 

\newcommand{\Tr}{\mathrm{Tr}} 

\newcommand{\trajc}[2]{\{X\}_{#1:#2}} 
\newcommand{\trajxv}[2]{\{\tilde{X}\}_{#1:#2}} 
\newcommand{\trajh}[2]{\{h\}_{#1:#2}}
\newcommand{\statec}[1]{X_{#1}} 
\newcommand{\statexv}[1]{\tilde{X}_{#1}}
\newcommand{\statexvh}[1]{(\tilde{X},h)_{#1}}
\newcommand{\SIlink}{Appendix}

\graphicspath{{./Figures/}}

\begin{document}

\title{Likelihood-based non-Markovian models from molecular dynamics}
\author{Hadrien Vroylandt}
\affiliation{Sorbonne Université, institut des sciences du calcul et des données, ISCD, F-75005 Paris, France}
\author{Ludovic Goudenège}
\affiliation{CNRS, FR 3487, Fédération de Mathématiques de
  CentraleSupélec, CentraleSupélec, 91190 Gif-sur-Yvette, Université Paris-Saclay, France}
\author{Pierre Monmarché}
\affiliation{Sorbonne Université,  Laboratoire Jacques-Louis Lions, LJLL, F-75005 Paris}
\affiliation{Sorbonne Université,  Laboratoire de Chimie Théorique, LCT, F-75005 Paris}
\author{Fabio Pietrucci} 
\affiliation{Sorbonne Université, Muséum National d’Histoire Naturelle, UMR CNRS 7590, Institut de Minéralogie, de Physique des
Matériaux et de Cosmochimie, IMPMC, F-75005 Paris, France}
\author{Benjamin Rotenberg}
\affiliation{Physicochimie des \'electrolytes et Nanosyst\`emes Interfaciaux, Sorbonne Universit\'e, CNRS, 4 Place Jussieu F-75005 Paris, France}

\date{\today}
\begin{abstract}
We introduce a new method to accurately and efficiently estimate the effective dynamics of collective variables in molecular simulations. Such reduced dynamics play an essential role in the study of a broad class of processes, ranging from chemical reactions in solution to conformational changes in biomolecules or phase transitions in condensed matter systems. The standard Markovian approximation often breaks down due to the lack of a proper separation of time scales and memory effects must be taken into account. Using a parametrization based on hidden auxiliary variables, we obtain a generalized Langevin equation by maximizing the statistical likelihood of the observed trajectories. Both the memory kernel and random noise are correctly recovered by this procedure. This data-driven approach provides a reduced dynamical model for multidimensional collective variables, enabling the accurate sampling of their long-time dynamical properties at a computational cost drastically reduced with respect to all-atom numerical simulations. The present strategy, based on the reproduction of the dynamics of trajectories rather than the memory kernel or the velocity-autocorrelation function, conveniently provides other observables beyond these two, including \textit{e.g.} stationary currents in non-equilibrium situations, or the distribution of first passage times between metastable states.
\end{abstract}

\maketitle

\section{Introduction and main results}
 
In different branches of Science, the interpretation and mathematical modeling of both experimental and computational data requires the analysis of the system dynamics in terms of a reduced set of collective variables (CVs), or order parameters. Prominent examples include chemical reactions in solution,  conformational changes in biomolecules or phase transitions in condensed matter systems. A standard approach is to approximate the evolution of the CVs by an effective dynamics, namely a closed equation in which the degrees of freedom beyond the CVs (forming the so-called environment or ``bath'') do not appear explicitly. Such coarse-grained models not only provide a physical interpretation more accessible to understanding than the full system, but also, from a numerical perspective, enable one to recover the desired dynamical properties with long but cheap simulations of the reduced system (while only shorter simulations of the large system are used to determine the effective dynamics).

The most widespread model for this task is the Langevin equation, which can be derived -- in some particular cases -- from the Hamiltonian dynamics of a small system interacting with a large environment. It describes the evolution of a Markov process, which requires that the decorrelation time of the environment is short compared to the characteristic times of the reduced system. However many cases do not enter the validity range of this approximation, displaying memory effects~\cite{Hynes1985,Bergsma1987,Bocquet1997,Daldrop2018,Min2005,Kheifets2014,Lysy2016,Mitterwallner2020a}. To go beyond the Markovian approximation, a popular class of processes is given by the generalized Langevin equation (GLE)~\cite{Zwanzig2001,Chorin2000,Chorin2002,Ma2016,Chung2019,Darve2009,Izvekov2013}
\begin{equation}
  \label{eq:GLE_mem}
  \begin{cases}
      \dot{x}(t)= v(t) \\
  M\dot{v}(t) = F_\text{eff}(x(t)) - \int_0^t K(t-\tau) v(\tau) \dd \tau +R(t),
  \end{cases}
\end{equation}
where $x(t)$ is the value of the $d$-dimensional collective variable at time $t$, $v(t)$ its time derivative, $M$ is an effective mass, $F_\text{eff}$ is a mean force, usually deriving from a potential $V$ identified with the free energy, $K$ a memory kernel and $R(t)$ a (colored) noise.

This form of the GLE can be motivated from the dynamics of the original full system following the Mori-Zwanzig formalism~\cite{Zwanzig1973,Mori1965,Zwanzig2001,Mori1965a}, even though it cannot be formally obtained as a controlled approximation of the exact coarse-grained dynamics, since a rigorous derivation generally results in a memory kernel that depends on the CVs~\cite{Glatzel2021,vroylandt2022arXiv}. Nevertheless, in practice this simple form is the most widely used effective dynamics.  While an analytical derivation of the memory kernel is possible only in a few cases~\cite{Doerries2021}, for more general systems, $K$ can be estimated from a data-driven approach. In most cases, the goal is to extract the memory kernel from trajectories of the CV computed with all-atom simulations~\cite{Berkowitz1981,Berne1970,Berne1990,Lei2016,Daldrop2018,Carof2014,Lesnicki2016,Jung2017,Jung2018,Klippenstein2021,Lei2010,Davtyan2015,Li2015,Li2017,Yoshimoto2017,Straube2020,Ayaz2021.}

As already mentioned, the solutions of \eqref{eq:GLE_mem} are not Markov processes, except when $K$ is the Dirac $\delta$ function and $R$ is a white noise. Both for fitting the model and then for generating new trajectories of the effective dynamics, it is convenient to consider the subclass of models where an extended process $(x,v,h)$ is Markovian, with  $h$ some hidden auxiliary variables \cite{Fricks2009,Lee2019,Ceriotti2010,Baczewski2013,Ciccotti1980,Stella2014, Ma2016,Wang2020,Bockius2021}. Restricting further to the case where the evolution of the hidden variables and the coupling with the observed variables are linear, this leads to an equation of the form
\begin{equation}
  \label{eq:aux_var_sys}
  \hspace{-0.26cm}
  \begin{cases}
    \dot x =  v\\
    \dot v = M^{-1}F_\text{eff}(x) - A_{vh} h  -A_{vv} v + \sigma_{vv} \xi(t)+ \sigma_{vh} W(t)\\
    \dot{h} = \hspace{1.68cm} -A_{hh} h - A_{hv} v +\sigma_{vh}^\T \xi(t)+ \sigma_{hh} W(t)
  \end{cases}
\end{equation}
where $A_{vh},A_{vv},A_{hh},A_{hv},\sigma_{vv},\sigma_{vh},\sigma_{hh}$ are constant matrices and $\xi$ and $W$ are independent standard white noises. This gives a convenient class of models parametrized by the dimension $d_h$ of $h$, the corresponding matrices and the (rescaled) effective force $M^{-1} F_\text{eff}$. For equilibrium processes, the coefficients of \eqref{eq:aux_var_sys} are related by the so-called Fluctuation-Dissipation relation \cite{Ceriotti2010}. Although we could enforce this condition, thereby reducing the number of parameters, we do not since we also consider non-equilibrium systems in the following.

Integrating over the hidden variables, we recover \eqref{eq:GLE_mem} with a memory kernel of the form of a finite Prony series \cite{Ceriotti2010,Baczewski2013}
\begin{equation}
    \label{eq:prony_series}
    K(\tau) =  w_0 \delta(\tau) + \sum_{k=1}^{d_h} w_k e^{-\lambda_k \tau}
\end{equation}
where $w_k$ and $\lambda_k$ are (possibly complex) coefficients of the series derived from the matrices $A_{vh},A_{vv},A_{hh},A_{hv}$.
In principle, on all finite time intervals, any kernel given as the sum of a Dirac function at zero and of a continuous function can be approximated arbitrarily accurately by a sum of the form \eqref{eq:prony_series}. However, in practice $d_h$ is relatively small and  memory kernel with e.g. algebraic tail can only be approximated on small time interval \cite{Fricks2009,Bockius2021}.

The use of auxiliary variables in the form of \eqref{eq:aux_var_sys} has been abundantly used and studied, as it allows efficient integration of GLE~\eqref{eq:GLE_mem} \cite{Ceriotti2010,Ciccotti1980,Ma2019a}, even though other methods exist~\cite{berkowitz1983,barrat2011,Jung2017,Jung2018,Li2015}.
The estimation of GLE parameters from simulations is an active field of research. The main method consists in a non-parametric estimation of the memory kernel via the Volterra integral equation \cite{Ayaz2021,Wang2019,Li2017,Lei2016,Ma2016,Ma2019a,Gottwald2015,Lee2019}, but other methods have also been proposed \cite{Bockius2021,Wang2020,Russo2019,Davtyan2015,Berne1990}.
In the present work, we i) introduce a novel parametric estimator of GLE coefficients, based on a maximum likelihood approach and ii) show that it allows building faithful coarse-grained models of MD simulations in a cost-effective way (i.e., starting from a relatively small training data set), such that the dynamics is well reproduced.

\section{Data-driven approach on extended dynamics}
\label{sec:data-driven-approch}

In statistics, a standard method to deal with hidden variables is the Expectation-Maximization (EM) algorithm, which belongs to the category of likelihood maximization algorithms \cite{Dempster1977,Little2019}. It is of frequent use to estimate parameters of time series models in the case of partial or noisy observations of the system, either for hidden Markov models \cite{Rabiner1989} or state-space models \cite{Dembo1986a}. A first application in the context of GLE was proposed in Ref.~\citenum{Fricks2009} to reconstruct the memory kernel in the absence of effective force $F_\text{eff}(x)$ and under more restrictive conditions than the method presented below. 

The algorithm proceeds by alternating steps: In the E-step, one determines the conditional probability law of the hidden variables given the observed ones at fixed parameters; in the M-step, one optimizes the parameters to maximize the log-likelihood averaged with respect to these conditional laws. In the following we denote as $\Theta_j$ the whole set of parameters estimated after $j$ iterations of the algorithm, which includes the mean force projected on some functional basis (which can be very large in general, or reduced if prior knowledge on the system is available), the coefficients of the matrices $A,\sigma$ of \eqref{eq:aux_var_sys} and, for technical reasons discussed below, the mean value at time zero of the hidden variables, $\avg{h_0}$.

\subsection{EM algorithm}
\label{sec:em-algorithm}
The available data, obtained from all-atom simulations, consists of a set of independent trajectories. For simplicity of the notation, we introduce the algorithm with only one trajectory $\{x\}_{0:N} = \{x({k\Delta t}), k\in \llbracket 0,N\rrbracket\}$  for some timestep $\Delta t$ and simulation time $T=N\Delta t$, the extension to the general case being straightforward. The statistical models we consider are Euler-Maruyama discretizations of \eqref{eq:aux_var_sys} with the same timestep $\Delta t$, for a fixed dimension $d_h$ of auxiliary variables $h$. The state of the system at time $t=k\Delta t$ will be denoted $(x_k,v_k,h_k) = \statexvh{k} = \statec{k} $  and we write $\trajc{0}{N}$ a complete trajectory of the system. Hence, $ \statexv{}$ is the value of the known variables since, from the choice of the Euler-Maruyama scheme,  the velocity can be computed as  $v_k=(x_{k+1}-x_k)/\Delta t $. 
In the following we write $\pi(z)$ the probability density of a variable $z$, $\pi(z|u)$ the conditional probability density of $z$ with respect to $u$ and, in both cases, $\pi_{\Theta}$ to explicit the value of the parameters if needed. 

As the extended system is Markovian, we have for the probability  density of a trajectory 
\begin{equation}
  \label{eq:pathProbhidden}
  \pi\left(\trajc{0}{N} \right) =  \prob{\statec{0}}
  \times \prod_{k=0}^{N-1} \pi{(\statec{k+1} | \statec{k}) }
\end{equation}
and the form of \eqref{eq:aux_var_sys} and of the Euler-Maruyama scheme lead to a Gaussian transition kernel, characterized by its mean $\mu$ and variance $\Sigma$ (see \SIlink).  

\paragraph{E-step}

\begin{figure}[ht]
  \centering
  \includegraphics[width=0.9\linewidth]{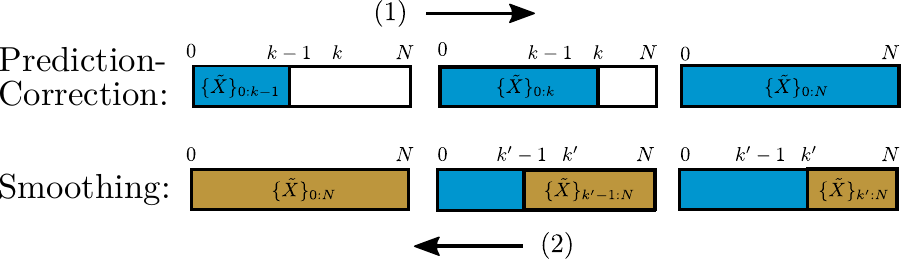}
  \caption{E-step. (1) We first predict iteratively the history of $h_t$ for the whole trajectory, using a predictor-corrector.
  (2) The values of $h_t$ are then smoothed iteratively backwards from the end of the trajectory (see text).
  \label{fig:filtersmoother}
  }
\end{figure}

The first step is to  compute the conditional law of the hidden variables
given the observed variables at the current guess of the parameters, i.e. 
$\pi_{\Theta_j}( \trajh{0}{N} |\trajxv{0}{N})$. Due to the Markovianity of the extended system, it is sufficient to compute the mean and variance of the Gaussian marginal laws $ \pi( h_k, h_{k+1}| \trajxv{0}{N})$ for all $k\in \llbracket 0,N-1\rrbracket$. Taking advantage of the explicit form of the transition probability (Eq.~(3) in the S.I.), we apply an iterative  
predictor-corrector-smoother approach (also known as Kalman filter and Rauch-Tung-Striebel smoother) \cite{Fildes1991}. 
Starting from the trajectory up to step $k-1$, we determine the law of the hidden variable $h_{k}$ conditioned on the past information $\trajxv{0}{k-1}$ only. We then use the expression of the  transition probability $ \pi{(\statec{k} | \statec{k-1}) }$ to determine the current value of $ \prob{h_{k}| \trajxv{0}{k}}$. These are the prediction and correction parts that are run forward on the trajectory, i.e. from $k=0$ to $k=N$ (arrow (1) on Fig.~\ref{fig:filtersmoother}). The initial guess at $k=0$ of $\prob{h_{0}}$ uses the measured $\avg{h_0}$ vector as the mean and an arbitrary variance (identity matrix). Such initial guess could be optimized, but we did not observe any influence on the final results. The second part of the E-step, called the smoother part,  computes $\pi(h_{k-1}|h_{k},\trajxv{0}{N})$ and is run backward, i.e. from $k=N$ to $k=0$ (arrow (2) on Fig.~\ref{fig:filtersmoother}) which finally gives the required probability law of $h_{k-1}, h_{k} $ conditioned to the full observed trajectory. Detailed formula are presented in \SIlink.

\paragraph{M-step}

For any set of parameters $\Theta$, introduce the evidence lower bound $\mathcal{L}^{j}_{LB}$ after the $j$-th iteration of the algorithm as the expectation with respect to $\pi_{\Theta_j}(\trajh{0}{N}|\trajxv{0}{N})$ of the log-likelihood of  the full trajectory $\trajc{0}{N}$ with parameter $\Theta$, namely (see derivation in \SIlink)
\begin{align}
\label{eq:evidencelowerbound}
    \mathcal{L}^{j}_{LB}(\Theta) = &\int   \pi_{\Theta_j}(\trajh{0}{N}|\trajxv{0}{N}) \ln \pi_\Theta\left(\trajc{0}{N}\right) \dd \trajh{0}{N}  \nonumber \\
    =&\int \pi_{\Theta_j}(h_{0}|\trajxv{0}{N}) \ln  \pi_\Theta{( \statec{0}) }  \dd h_0 \nonumber \\
    &+\sum_{k=0}^{N-1} \int  \pi_{\Theta_j}(h_{k}, h_{k+1}|\trajxv{0}{N})  \nonumber\\
   & \hspace{2cm} \times \ln  \pi_\Theta{(\statec{k+1} | \statec{k}) } \dd h_k \dd h_{k+1}.
\end{align}
The M-step consists in setting $\Theta_{j+1}$ to be the maximizer of this quantity. Notice that due to the particular form of \eqref{eq:aux_var_sys}, $\mathcal{L}^{j}_{LB}(\Theta) $ is an explicit function of $\Theta$, that can be easily optimised as described in \SIlink.

\paragraph{Full algorithm}

The algorithm then run as follows. An initial random or informed guess $\Theta_{0} $ is taken for the parameters. Such informed guess could come from a previous execution of the algorithm with a different number of hidden dimensions. From parameters $\Theta_{j} $, a new set of parameters $\Theta_{j+1} $ is computed through an iteration of E and M steps. Since maximizing the evidence lower bound $\mathcal{L}^{j}_{LB}$ increases the observed likelihood, the method is iterated until either a prescribed maximum number of EM steps or a convergence criterion is reached.

\paragraph{Assessing the quality of a given model}
\label{sec:model-validation}

The number of hidden dimensions $d_h$ is an important parameter of the algorithm. It can be chosen using a model validation approach, classically by dividing the set of trajectories between a training and a validation set. However here, we simply compute the optimal parameters for several values of $d_h$ and compare the predictions of the corresponding models for a number of observable properties, such as the memory kernel, velocity-autocorrelation functions (VACF) or mean first passage times.
Similarly, the quality of the model depends on the time step used for the coarse-grained dynamics. This choice depends among other things on the numerical scheme for the propagator. For a given underlying dynamics of the full system, the most accurate choice for the coarse-grained one is to use the same time step $\Delta t_{full}$, but as a compromise with the amount of data one can also use $\Delta t = m\Delta t_{full}$ (\textit{i.e.} using only every $m$ step), with $m$ a small integer.

\paragraph{Efficient sampling of new trajectories}

Once the model has been optimized by the EM algorithm, it can be used to generate new trajectories in the CV-space. Due to their limited computational cost compared to MD trajectories, such synthetic data grants easier access to well-converged average properties, in the form of static and dynamic observables. As an example, in section \ref{sec:results} the mean first passage times (as well as their probability densities) of a Lennard-Jones dimer in a bath are estimated based on the GLE model and compared with the corresponding ones extracted from expensive MD simulations.

\section{Results}
\label{sec:results}
\begin{figure}[ht]
  \centering
  \includegraphics[width=\linewidth]{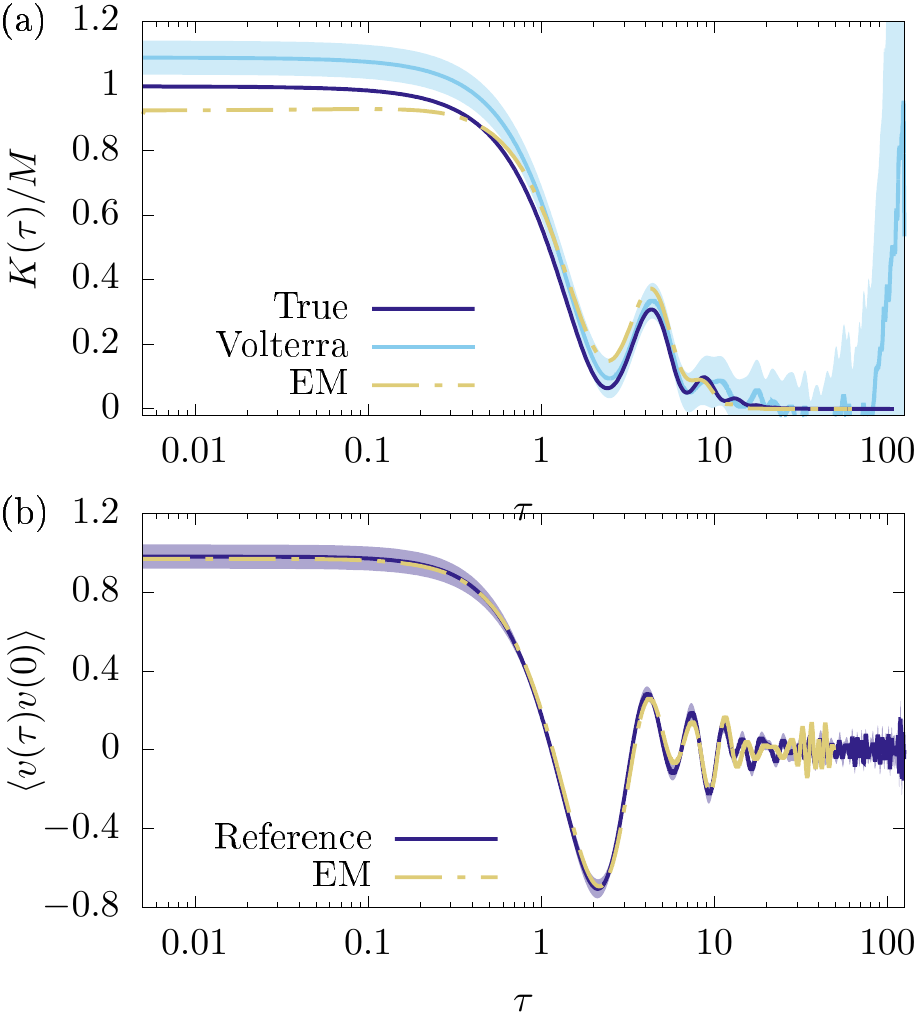}
  \caption{
  Equilibrium 1D case.
  (a) Memory kernel $K(\tau)$ divided by the mass $M$: The true kernel used to generate the reference trajectories (dark blue line) is compared with the predictions of the Volterra method (cyan solid line, with shaded area indicating uncertainties computed from a bootstrap analysis) and of the present EM method (dashed-dotted yellow line).
  (b) Velocity autocorrelation function, from the reference trajectories (dark blue line) and from new trajectories sampled using the fitted EM model (dashed-dotted yellow line).}
  \label{fig:lm}
\end{figure}

We first present the result of the algorithm on a simple yet non-trivial test case with a 1D system following the extended dynamics of \eqref{eq:aux_var_sys}, with $5$ hidden dimensions and a quadratic potential well $V(x)=x^2/2$, using $20$ trajectories of $2.5\cdot 10^{4}$ steps and a timestep of $5\cdot 10^{-3}$. The effective force is fitted as a linear function of $x$. Fig.~\ref{fig:lm}a compares the result of our algorithm to the true memory function that can be computed from \eqref{eq:prony_series} and the one obtained by the Volterra method (see Materials and Methods). It demonstrates that the present EM method is able to reproduce the true memory kernel. Furthermore, the parametric structure of the fitted model enforces the decay to zero of the memory kernel, whereas the Volterra method is unstable at long time. Fig.~\ref{fig:lm}b finally shows that the method accurately reproduces the VACF.

\begin{figure*}[ht]
  \centering
  \includegraphics[width=\linewidth]{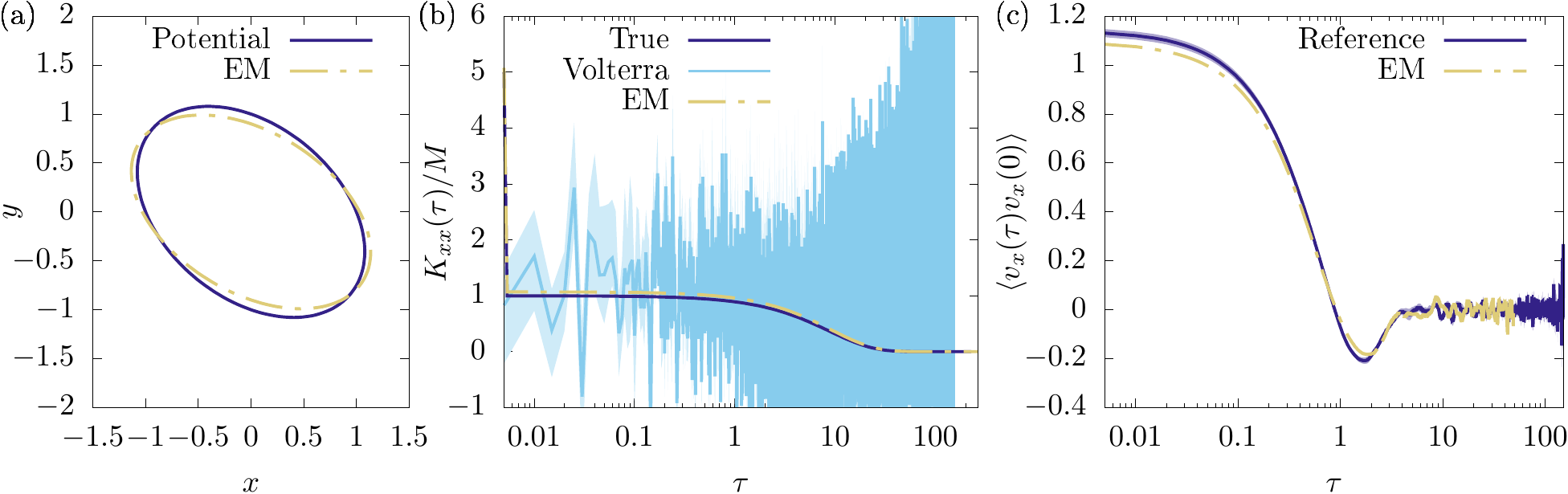}
  \caption{
  Non-equilibrium 2D case. (a) Locus of $V(x,y)=1$ for the original potential and the one estimated by the EM algorithm. (b) $xx$ component of the reconstructed memory kernel $K_{xx}(\tau)$ divided by the mass $M$: The true kernel used to generate the reference trajectories (dark blue line) is compared with the predictions of the Volterra method (cyan solid line, with shaded area indicating uncertainties computed from a bootstrap analysis) and of the present EM method (dashed-dotted yellow line); the left peak represents the Dirac function of \eqref{eq:prony_series}. (c) Velocity autocorrelation function (for the $x$ component of the velocity), from the reference trajectories (dark blue line) and from new trajectories sampled using the fitted EM model (dashed-dotted yellow line).}
  \label{fig:2d}
\end{figure*}

The algorithm also applies to multidimensional and nonequilibrium systems. This is illustrated on Fig.~\ref{fig:2d} for a 2D system with two different thermal noises along each axis, with temperatures $T_x=1$ and $T_y=5$ and a quadratic potential $V(x,y)=\frac{1}{2}\left(x^2+\frac{3}{4}xy+y^2\right)$ whose principal axes are not aligned with the $x$ and $y$ axes, leading to non-equilibrium conditions. This setup is inspired by a similar Markovian model used to describe non-equilibrium experiments on cold atoms \cite{Mancois2018}. We run $20$ trajectories of  $3\cdot 10^{4}$ steps with a timestep of $5\cdot10^{-3}$. The effective 2D force is fitted as a linear combination of $x$ and $y$. The corresponding quadratic potential, illustrated in Fig.~\ref{fig:2d}a, is in good agreement with the one used to generate the trajectories. Fig.~\ref{fig:2d}b then shows that the algorithm correctly estimates the memory kernel (in the present case, a simple one with a single hidden dimension for each visible dimension). In particular, the presence of strong Markovian component is captured by the algorithm but missed by the Volterra method. Finally, the dynamics of the system is well reproduced, as demonstrated for the VACF on Fig.~\ref{fig:2d}c. 

\begin{figure}[ht]
  \centering
  \includegraphics{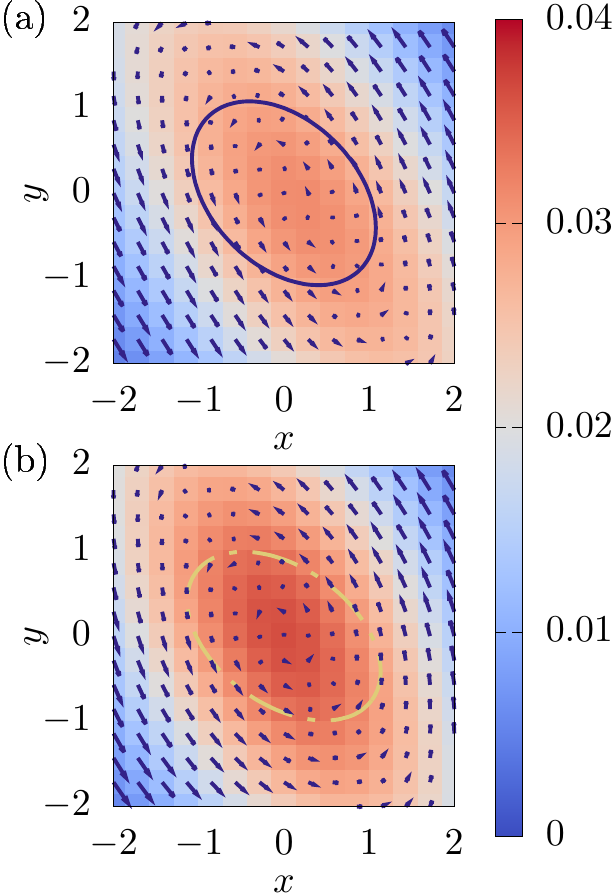}
  \caption{Non-equilibrium 2D case: beyond the kernel and the VACF. Stationary probability distribution (colors) and average velocity (arrows) as function of the position for (a) the original dynamics and (b) the GLE model estimated by the EM algorithm. The two ellipses are the same as in Fig.~\ref{fig:2d}a and represent the locus of $V(x,y)=1$ for the original potential (blue line) and the one estimated by the EM algorithm (dashed yellow line).
  }
  \label{fig:2d_vect}
\end{figure}

The present approach, based on the reproduction of the dynamics of trajectories rather than the memory kernel or the VACF, conveniently provides other observables beyond these two. Indeed, by generating new trajectories corresponding to the fitted GLE model, one has in principle access to all properties that can be computed from the time evolution of the collective variables. As an illustration, Fig.~\ref{fig:2d_vect} shows for the same non-equilibrium 2D case the stationary probability distribution and the average velocity as a function of the position, estimated using either the initial trajectories used to fit the GLE model (panel~\ref{fig:2d_vect}a) or the same number of trajectories generated with the latter (panel~\ref{fig:2d_vect}b). Despite the relatively small number (only 20) of original trajectories used to fit the model and to compute the properties, those computed from the fitted GLE model are in very good agreement with the original ones.

As a final illustration, we apply our algorithm to a more realistic 3D system composed of $512$ Lennard-Jones (LJ) particles at reduced temperature $\hat{T}=k_BT/\epsilon =1$ and reduced density $\hat \rho = \rho\sigma^3= 1$. Two of the LJ particles are singled out to form a dimer \cite{Berne1990}, the others constituting the solvent. The CV of interest is the distance $r$ between the two particles forming the dimer. LJ parameters for all interactions are taken as $\epsilon=1$ and $\sigma=1$ (in LJ units), except between the two particles forming the dimer, with $\epsilon_d=2$ and $\sigma=1$. The size of the cubic simulation box is $8\sigma$, with periodic boundary conditions in all directions. The dynamics is integrated with a time step of $\Delta t_{MD}= 10^{-3}$ (in LJ units) in the NVE ensemble using the LAMMPS simulation package \cite{Plimpton1995}. We run 20 trajectories with length of $10^6 $ timesteps and CV values are extracted every 2 steps.

We fit GLE models \eqref{eq:aux_var_sys} with the EM algorithm for a number of hidden dimensions ranging from 3 to 6. In all cases, the effective force $F\e{eff}(r)$ determined from the MD trajectory is used as the single function of the above-mentioned functional basis, so that fitting this part reduces to determining a single prefactor. Our aim is to test the ability of these models to reproduce, in the statistical sense, the properties of the original simulations. In order to check the importance of the hidden variables, we also provide an analysis for a Markovian model, fitted using a maximum likelihood algorithm with $0$ hidden dimensions (corresponding to the M-step of the above EM algorithm). For each fitted GLE model, we generate $75$ new trajectories of length $10^5$ timesteps using \eqref{eq:aux_var_sys}, to compute the observable properties and compare them with those obtained from the original set of MD trajectories. 

\begin{figure*}[ht]
  \centering
   \includegraphics[width=\linewidth]{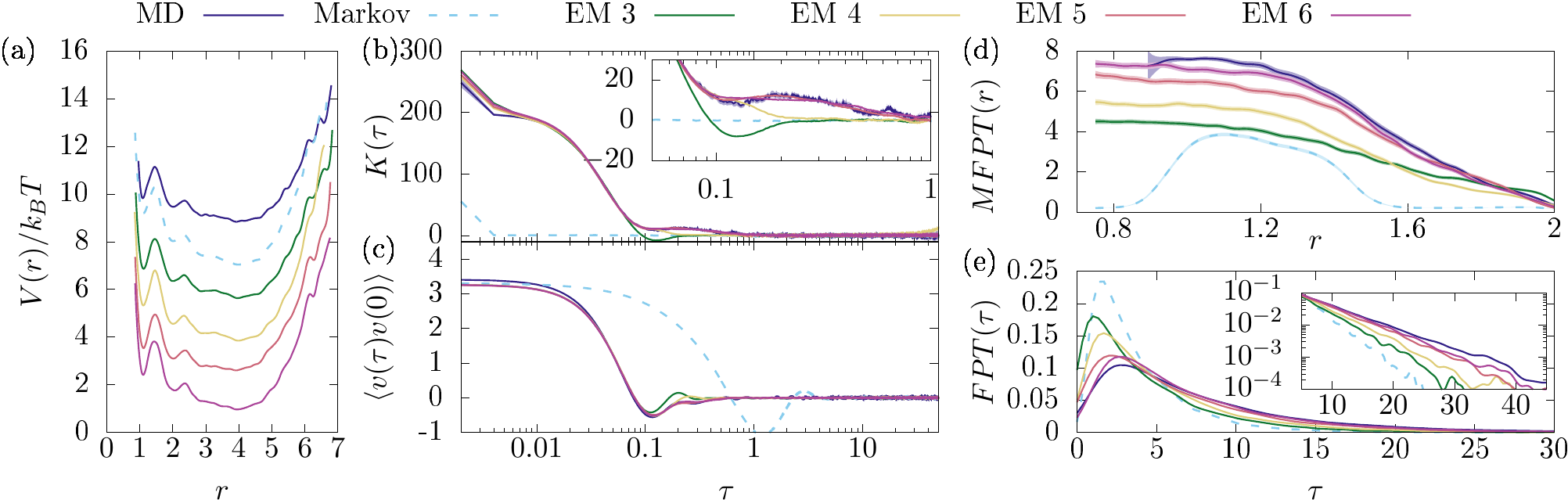}
  \caption{
  Lennard-Jones fluid: two solutes in an explicit solvent. In all panels, results are shown for the reference MD trajectory (dark blue line) and for trajectories generated by the estimated Markov model (dashed cyan lines) and EM models with 3 to 6 hidden dimensions (from dark green to purple). Unless specified, all quantities are in LJ units.
  (a) Free energy (in units of the thermal energy, $k_BT$) estimated from histograms of the distance $r$ between the two solutes; the various cases are shifted by physically irrelevant constants for clarity. 
  (b) Memory kernel compared and (c) velocity autocorrelation function; the inset in panel (b) shows a zoom on intermediate times.
  (d) Mean first passage time to reach $r=2.0$ starting from the distance $r$. 
  (e) Distribution of the first passage time for trajectories starting at $r=2^{1/6}$ and ending at $r=2.0$ (the inset shows a zoom on the tail of the distributions, on semi-logarithmic scale).
  \label{fig:dimer}}
\end{figure*}

We first compare the stationary distribution for the various GLE models in Fig.~\ref{fig:dimer}a, which shows the free energy as a function of the $r$ coordinate computed from the histogram of each set of new trajectories (\textit{i.e.} not the one corresponding to the fitted effective force). The good agreement with the MD free energy profile demonstrates (i) that the coefficient multiplying the model free energy profile of each model is fitted precisely and (ii) that the numerical integration of the GLE models is performed accurately. Notice that free energy beyond $r=4\sigma$ is affected by the size of the periodic box. The free energy displays two potential wells at $r=1.12\sigma$ and $r=2.00\sigma$, corresponding to the contact pair (CP, \textit{i.e.} the dimer) and the solvent shared pair (SSP, with solvent atoms belonging to the solvation shells of both solutes), whose dynamics is investigated below.

We then consider dynamical observables in Fig.~\ref{fig:dimer}b, which shows the memory kernel estimated from the Volterra method \cite{Daldrop2018} for MD as well as GLE trajectories, and Fig.~\ref{fig:dimer}c, which illustrates the VACF (the velocities being computed numerically from positions both in the MD and GLE trajectories). In both cases, increasing the number of hidden dimensions increases the fidelity of the model with respect to the original data. The latter are correctly reproduced for $5$ and $6$ hidden dimensions. The plot also shows the poor quality of the Markovian model, which confirms the necessity of introducing some hidden variables \cite{Ayaz2021}.

Finally, we study the transition kinetics between the CP and SSP states,
as a stringent test requiring accurate reproduction of both thermodynamic and dynamic properties of the system. Fig.~\ref{fig:dimer}d represents the mean first passage time (FPT) to reach the SSP state starting from  smaller $r$ distances, whereas  Fig.~\ref{fig:dimer}e represents the FPT distribution for trajectories starting from the CP state and reaching the SSP state. Clearly, a sufficient number of hidden dimensions (in this case 5-6) allows to quantitatively reproduce the detailed transition statistics. This demonstrates again both the importance of memory effects and the ability of the present algorithm to reconstruct an accurate GLE model.

\section*{Conclusions}
\label{sec:conclusion}

In this work we addressed the construction of reduced mathematical models of the dynamics of complex molecular systems. 
Projecting the phase-space trajectories on a reduced set of collective variables leads to a powerful framework for the prediction of thermodynamic and kinetic properties of experimental interest. 
However, the key problems in this context consist in the identification of a suitable dynamical equation and its parametrization.
We developed a novel approach combining generalized Langevin equations, their numerically-efficient representation via Markovian equations including hidden variables, and a powerful 
machine-learning algorithm borrowed from the field of statistical modeling and data science.
Starting from non-Markovian trajectories (\eg projected all-atom molecular dynamics trajectories in condensed-matter applications), we maximize the likelihood of an extended Markovian model employing the expectation-maximization algorithm. The advantage of obtaining an explicit parametrization allows for inexpensive sampling of synthetic trajectories, that can be used for the direct computation of quantitative observables (beyond the standard memory kernel and VACF) such as stationary currents in non-equilibrium situations, or the distribution of first passage times between metastable states, generally hard to access through atomistic simulations. 

Several features distinguish our approach from others existing in the literature. Firstly, the model we optimize includes an explicit parametrization of both the friction and the noise, ensuring consistency between the analysis of the MD trajectories and the generation of new projected trajectories. Secondly, our method is based on a maximum likelihood procedure, which is well justified from a mathematical perspective. In particular, instead of estimating a non-parametric kernel which is then parametrized (as \eg in Volterra-based approaches), the parametric model is directly fitted on the data; this should limit the accumulation of errors. Thirdly, we do not enforce equilibrium conditions (such as the fluctuation dissipation theorem) on the model, so that the present approach offers the possibility to investigate non-equilibrium systems.  Finally, the present approach readily applies to multidimensional CVs and corresponding matrix memory kernels.

The maximum likelihood approach offers a versatile strategy to implement various extended Markovian models, which could be extended in particular to position-dependent generalized Langevin equations and higher order discretization schemes. Overall, the present work provides an efficient way to generate reduced dynamical models for multidimensional collective variables, with the same memory kernels as the underlying complex system, enabling the accurate sampling of the long-time dynamics of the latter at a dramatically reduced computational cost.

\begin{acknowledgments}
  We thank Sara Bonella, Michele Casula, Arthur France-Lanord, Marco Saitta, Mathieu Salanne, and Rodolphe Vuilleumier for fruitful discussions within the MAESTRO collaboration.
This project received funding from the European Research Council under the European Union's Horizon 2020 research and innovation program (Grant Agreement No.~863473).
\end{acknowledgments}

\section*{Authors contributions}
\label{sec:auth-contr}

H.V. implemented the new algorithm and performed simulations. H.V., L.G., P.M, F.P. and B.R. developed the methodology, analyzed the results and wrote the manuscript.

\section*{Materials and methods}
\paragraph{Estimate of the (potential of) mean force}
In the first two examples, the coefficients of the quadratic potentials in the EM method follow from those of the corresponding forces, which are the ones determined numerically along with the parameters related to the memory (see~\SIlink). Potentials of mean force in the Volterra method result from quadratic fits of the logarithm of histograms of the position. We obtain the results for the memory kernel with the Volterra approach, using the memtools package (\url{https://github.com/jandaldrop/memtools}~\cite{Daldrop2018}) in the 1D case and the multidimensional version of Ref.~\citenum{Lee2019} (see the link to our implementation below) in the 2D case.

\paragraph{MD simulation details for the LJ dimer}  The dynamics is integrated with a time step of $\Delta t_{MD}= 0.001$ (in LJ units) in the NVE ensemble with the velocity-Verlet algorithm using the LAMMPS simulation package~\cite{Plimpton1995}. We run 20 trajectories of $10^6$ timesteps and CV values are extracted every 2 steps.

\paragraph{EM convergence} Initial values of the parameters are taken randomly. For all examples, we stop the EM iterations if the difference of log-likelihood between two EM steps is less than $10^{-8}$ or if the number of EM steps exceeds $2000$. 

\paragraph{Density and average velocity for the non-equilibrium 2D case}  The density is estimated by kernel density estimation using the positions along the trajectories. The average velocities are estimated conditionally on the positions using kernel regression. The same Gaussian kernel is used in both cases, with a bandwidth of $1$. For Fig.~\ref{fig:2d_vect}b, $20$ new trajectories of $3\cdot 10^{4}$ steps with a timestep of $5\cdot10^{-3}$ were sampled from the fitted GLE model and compared to the original 20 trajectories of Fig.~\ref{fig:2d_vect}a.

\paragraph{Mean first passage time estimation}
The FPT is estimated for molecular dynamics starting by restraining the initial position with a parabolic potential as a function of $r$ using PLUMED~\cite{plumed}. $2000$ trajectories are generated from different restrained positions. Gaussian kernel estimates (with bandwidth of $1/\Delta t_{MD}$) of the mean FPT as well as the FPT density are then obtained conditioned on the realized starting position. The FPT and MFPT from the fitted models are computed using $1500$ trajectories per initial value of the distance, again employing kernel estimates.

\paragraph{Code availability}

A python package to perform the analysis introduced in the present work is available at \url{https://github.com/HadrienNU/GLE_AnalysisEM}.
Our implementation of the 2D Volterra method is available here: \url{https://github.com/HadrienNU/VolterraBasis}


\appendix

\begin{widetext}

\section*{Appendix}
  
\subsection*{Likelihood of a trajectory} 
We introduce the Euler-Maruyama discretizations of Eq.~(2) in the main text
\begin{equation}
  \label{eq:euler_scheme}
  \begin{cases}
  x_{k+1} =  x_k + v_k \Delta t\\
  v_{k+1}= v_k + \Delta t\left(M^{-1} F_\text{eff}(x_{k})  - A_{vv} v_k - A_{vh}h_{k}\right) +\sqrt{\Delta t}\sigma_{vv} G_v +\sqrt{\Delta t}\sigma_{vh} G_h   \\
 h_{k+1} =  h_{k}  - \Delta t \left(A_{hv} v_k + A_{hh} h_{k} \right)  +\sqrt{\Delta t}\sigma_{vh}^\T G_v +\sqrt{\Delta t} \sigma_{hh} G_h 
  \end{cases}
\end{equation}
where $G_v$ and $G_h$ are random centered reduced Gaussian vectors. To alleviate notations, we introduce the vector $\mu_{\Delta t}(\statec{k})$ and the matrix $\sigma_{\Delta t}$ such that the last two equations of  \eqref{eq:euler_scheme} read
\begin{equation}
\label{eq:euler_random_part}
    \begin{pmatrix} v_{k+1} \\ h_{k+1}\end{pmatrix} = \mu_{\Delta t}(\statec{k}) + \sigma_{\Delta t}   \begin{pmatrix} G_v \\ G_h\end{pmatrix}.
\end{equation}
The transition probability density in Eq.~(4) of the main text is
\begin{equation}
    \label{eq:transprob_app}
  \pi{(\statec{k+1} | \statec{k}) }  = \mathcal{N}\left(\begin{pmatrix} v_{k+1} \\ h_{k+1}\end{pmatrix}
    ; \mu_{\Delta t}(\statec{k}) , \sigma_{\Delta t} \sigma_{\Delta t}^\T\right) \delta\left( v_k-\f{x_{k+1}-x_k}{\Delta t} \right)
\end{equation}
where
\begin{equation*}
    \mathcal{N}\left(y ;\mu,\Sigma\right)= \f{e^{-\f{1}{2}\left[y-\mu \right]^\T \Sigma^{-1}\left[y-\mu\right]}}{\sqrt{(2\pi)^{D} \det \Sigma }} =   \f{e^{-\f{1}{2} \Tr \left[ \Sigma^{-1}\left[y-\mu\right]\left[y-\mu \right]^\T \right]}}{\sqrt{(2\pi)^{D} \det \Sigma }} 
\end{equation*}
is a notation for a ($D$-variate) Gaussian distribution density for the variable $y$ with mean $\mu$ and variance $\Sigma$. Since the presence of the Dirac function imposes the velocity as $v_k=(x_{k+1}-x_k)/\Delta t $, we always assume this condition to be satisfied and consider in the following only the non-degenerate part of the transition probability. 

From Eq.~(4) in the main text, the log-likelihood of a trajectory is then given by
\begin{align*}
\label{eq:likelihood_traj}
   \ln   \pi_\Theta\left(\trajc{0}{N} \right) = \ln  \pi{(\statec{0})} - \f{1}{2}N\ln\left[ (2\pi)^{d+d_h} \det\sigma_{\Delta t}\sigma_{\Delta t}^\T\right]-\f{1}{2} \sum_{k=0}^{N-1} \Tr\left[ {(\sigma_{\Delta t}\sigma_{\Delta t}^\T)^{-1} \left( \statec{k+1}-\mu_{\Delta t}(\statec{k})\right)\left( \statec{k+1}-\mu_{\Delta t}(\statec{k})\right)^\T} \right].
\end{align*}

\subsection*{M-step} Our ultimate objective is to maximize, with respect to the parameters $\Theta= \{F_\text{eff},\,A_{vv},\,A_{vh},\,A_{hv},\,A_{hh},\,\sigma_{vv},\,\sigma_{vh},\,\sigma_{hh},\, \avg{h_0}\}$, the log-likelihood of the observed trajectory $\trajxv{0}{N}$  given by
\begin{equation*}
\label{eq:likelihood_traj}
   \ln \mathcal{L}(\Theta) =\ln\int \pi_{\Theta}(\trajc{0}{N})\dd \trajh{0}{N}=  \ln \int   \pi_{\Theta}(\trajh{0}{N}|\trajxv{0}{N})\pi_\Theta\left(\trajxv{0}{N}\right) \dd \trajh{0}{N} . 
\end{equation*}
However, there is no practical way to maximize this expression directly, due to the integration with respect to the hidden variables, so that the EM algorithm relies instead on another quantity. Given $\Theta_j$ the current guess of the parameters at the $j^{th}$ iteration of the EM algorithm, the evidence lower bound (Eq.~(5) in the main text) is defined by
\begin{align*}
\label{eq:evidencelowerbound_app}
    \mathcal{L}^{j}_{LB}(\Theta) = &\int   \pi_{\Theta_j}(\trajh{0}{N}|\trajxv{0}{N}) \ln \pi_\Theta\left(\trajc{0}{N}\right) \dd \trajh{0}{N}  \nonumber \\
    =& \int \pi_{\Theta_j}(h_{0}|\trajxv{0}{N}) \ln  \pi_\Theta{( \statec{0}) }  \dd h_0+ \sum_{k=0}^{N-1} \int  \pi_{\Theta_j}(h_{k}, h_{k+1}|\trajxv{0}{N}) \times \ln  \pi_\Theta{(\statec{k+1} | \statec{k}) } \dd h_k \dd h_{k+1} .
\end{align*}
Using a convexity inequality, it can be shown (see Section~8.4.1 of Ref.~\citenum{Little2019}) that, for all $\Theta$,
\[\ln \mathcal L(\Theta) - \ln \mathcal L(\Theta_j) \geqslant \mathcal{L}^{j}_{LB}(\Theta) - \mathcal{L}^{j}_{LB}(\Theta_j)\,. \]
The M step then consists in taking $\Theta_{j+1}$ as the maximizer of $\mathcal{L}^{j}_{LB}$, which ensures that $\ln \mathcal L(\Theta_{j+1}) \geqslant \ln \mathcal L(\Theta_j)$, \textit{i.e.} an increase of the log-likelihood at each iteration. 

Contrary to the log-likelihood, the evidence lower bound can be optimized in practice. Indeed, from the transition probability \eqref{eq:transprob_app}, the evidence lower bound is
\begin{align}
    \label{eq:evidence_lower_bound_expr}
     \mathcal{L}^{j}_{LB}(\Theta) =&  \ln  \pi{(\statec{0})} - \f{1}{2}N\ln\left[ (2\pi)^{d+d_h} \det\sigma_{\Delta t}\sigma_{\Delta t}^\T\right] \nonumber \\
     &-\f{1}{2} \sum_{k=0}^{N-1} \Tr\left[ (\sigma_{\Delta t}\sigma_{\Delta t}^\T)^{-1} \left(\avg{\begin{pmatrix} v_{k+1} \\ h_{k+1}\end{pmatrix}\begin{pmatrix} v_{k+1} \\ h_{k+1}\end{pmatrix}^\T}_{\Theta_j} - \avg{\begin{pmatrix} v_{k+1} \\ h_{k+1}\end{pmatrix}\mu_{\Delta t}(\statec{k})^\T}_{\Theta_j} \right. \right. \nonumber \\
     &\left. \left. -\avg{\mu_{\Delta t}(\statec{k}) \begin{pmatrix} v_{k+1} \\ h_{k+1}\end{pmatrix}^\T}_{\Theta_j} +  \avg{ \mu_{\Delta t}(\statec{k})\mu_{\Delta t}(\statec{k})^\T}_{\Theta_j}\right) \right]
\end{align}
where all averages are with respect to $\pi_{\Theta_j}(h_{k}, h_{k+1}|\trajxv{0}{N})$. 
Let $B$ be the $d\times N_F$ matrix of the coefficients of the force in the functional basis, \textit{i.e.} $-M^{-1} F_\text{eff}(x) = B G(x)$ where $G=(G_1,\dots,G_{N_F})$ are the basis functions. From \eqref{eq:euler_scheme}, the vector $\mu_{\Delta t}(\statec{k})$ has a linear dependency in the parameters $(B,A_{vv},A_{vh},A_{hv},A_{hh})$ as it can be written as
\begin{equation*}
    \mu_{\Delta t}(\statec{k}) =\begin{pmatrix} v_{k} \\ h_{k}\end{pmatrix} -\Delta t \begin{pmatrix}
    A_{vv} & A_{vh} & B \\
    A_{hv} & A_{hh} & 0
    \end{pmatrix} \begin{pmatrix}
    v_k \\ h_k \\ G(x_k)
    \end{pmatrix}
    :=\begin{pmatrix} v_{k} \\ h_{k}\end{pmatrix} -\Delta t  \tilde\Theta\, \begin{pmatrix}
    v_k \\ h_k \\ G(x_k)
    \end{pmatrix}
\end{equation*}
where we have introduced a $(d+d_h) \times (d+d_h+N_F) $ matrix $ \tilde\Theta$.
As a consequence, the evidence lower bound reads
\[\mathcal{L}^{j}_{LB}(\Theta) = \ln  \pi{(\statec{0})} - \f{1}{2}N\ln\left[ (2\pi)^{d+d_h} \det\sigma_{\Delta t}\sigma_{\Delta t}^\T\right] -\Tr\left[(\sigma_{\Delta t}\sigma_{\Delta t}^\T)^{-1}\left( C_1 - \tilde \Theta C_2^\T- C_2\tilde \Theta^\T +\tilde  \Theta C_3 \tilde  \Theta^T  \right)\right]\]
where $C_1,C_2,C_3$ are matrices, independent from $\Theta$, which can be explicitly computed using \eqref{eq:evidence_lower_bound_expr} from the observations $\trajxv{0}{N}$ and the mean and covariance matrix of $\pi_{\Theta_j}(h_{k}, h_{k+1}|\trajxv{0}{N})$ for all $k$ (see the E-step below). The equation $\nabla_{\Theta}\mathcal{L}^{j}_{LB}(\Theta)=0$ can then be solved explicitly~\cite{Horenko2008,Espanol2011} and has the following unique  solution:
\begin{eqnarray*}
\tilde\Theta &=&   C_2^\T C_3^{-1}  \\
\sigma_{\Delta t}\sigma_{\Delta t}^\T & =& C_1 - \tilde \Theta C_2^\T- C_2\tilde \Theta^\T +\tilde  \Theta C_3 \tilde  \Theta^T\,.
\end{eqnarray*}

\subsection*{E-step}

The goal of the E-step is to compute $\pi_{\Theta_j}(h_{k}, h_{k+1}|\trajxv{0}{N})$, as required in the M-step, for a fixed $\Theta_j$. In the following we drop the subscript $\Theta_j$ and simply write $\pi=\pi_{\Theta_j}$.

First, the prediction-correction part of the E-step computes the probability distribution of hidden variables conditioned on the past trajectory of the visible variables, namely $\pi(h_k| \trajxv{0}{k})$, for all $k$. This is done  iteratively, forwards (\textit{i.e.} from $k=1$ to $k=N$),  using that
\begin{align}
  \label{eq:forwardterm2}
  \prob{h_k| \trajxv{0}{k}}=   \f{\int  \prob{h_{k-1} |\trajxv{0}{k-1}}  \pi{\left(\statexvh{k} | \statexvh{k-1}\right) \dd h_{k-1}}   }{ \prob{\statexv{k} | \trajxv{0}{k-1}}}.
\end{align}
Since the denominator does not depend on $h_k$ and, for all $k$, $\pi(\statexvh{k} | \statexvh{k-1})$ is a Gaussian density, it follows that $\pi (h_k| \trajxv{0}{k})$ is a Gaussian distribution for all $k$ and that its mean and covariance matrix can be computed by induction on $k$ (see~\eqref{eq:filter_res} below for the explicit expression).

This first part is followed by the Rauch-Tung-Striebel smoother part of the E-step, where $\pi_{\Theta_j}(h_{k}, h_{k+1}|\trajxv{0}{N})$ is computed for all $k$, iteratively, backwards (\textit{i.e.} from $k=N-1$ to $k=0$), using the relation
\begin{equation}
  \label{eq:pair_estimate_smoother}
   \prob{h_k,h_{k+1}  | \trajxv{0}{N}} =   \prob{h_k | h_{k+1}, \trajxv{0}{N}}  \prob{h_{k+1} | \trajxv{0}{N}} .
\end{equation}
The second term can be computed iteratively starting with $k=N-1$, already computed in the prediction-correction part (for $k=N$ in Eq.~\ref{eq:forwardterm2}), as the marginal of the previous iteration (from time step $k+2$ to $k+1$)
\begin{align*}
  \label{eq:RTS_smoother}
  \prob{h_{k+1} | \trajxv{0}{N}} = \int  \prob{h_{k+1},h_{k+2}  | \trajxv{0}{N}} \dd h_{k+2} .  
\end{align*}
The first term of the right hand side of \eqref{eq:pair_estimate_smoother} is obtained from the results of the prediction-correction part  and the transition probability distribution, using that
\begin{align*}
  \prob{h_k | h_{k+1}, \trajxv{0}{k}}  &= \f{\prob{\statexvh{k+1}| h_{k}, \trajxv{0}{k}}  \prob{h_{k} | \trajxv{0}{k}}}{\int  \prob{\statexvh{k+1}| h_{k}, \trajxv{0}{k}}  \prob{h_{k} | \trajxv{0}{k}} \dd h_{k} }  = \f{  \pi{\left(\statexvh{k+1} | \statexvh{k}\right) }  \prob{h_{k} | \trajxv{0}{k}}}{\int   \pi{\left(\statexvh{k+1} | \statexvh{k}\right) }   \prob{h_{k} | \trajxv{0}{k}} \dd h_{k} } .
\end{align*}
Similarly to the prediction-correction part, these relations imply that $\pi_{\Theta_j}(h_{k}, h_{k+1}|\trajxv{0}{N})$ is a Gaussian distribution and that its mean and average can be computed by an induction relation from $k=N-1$ to $k=0$ (see~\eqref{eq:smoother_res} below for explicit expressions). This concludes the E-step.

The E-step  is illustrated in Fig.~\ref{fig:Estep}: here, we sample a trajectory $\trajc{0}{N}$ with known parameters $\Theta$ (with a single auxiliary variable, \textit{i.e.} $d_h=1$), and our goal is to reconstruct the law $\pi_{\Theta}(\trajh{0}{N}|\trajxv{0}{N} )$ of the trajectory of the hidden variable, using only the trajectory of the observed variables $\trajxv{0}{N}$. This conditional law, represented in Fig.~\ref{fig:Estep} by its mean $k\mapsto \avg{h_k}$ and
twice its standard deviation (blue area), is concentrated on the original realization.

\begin{figure}[ht]
  \centering
  
  \includegraphics[width=0.9\linewidth]{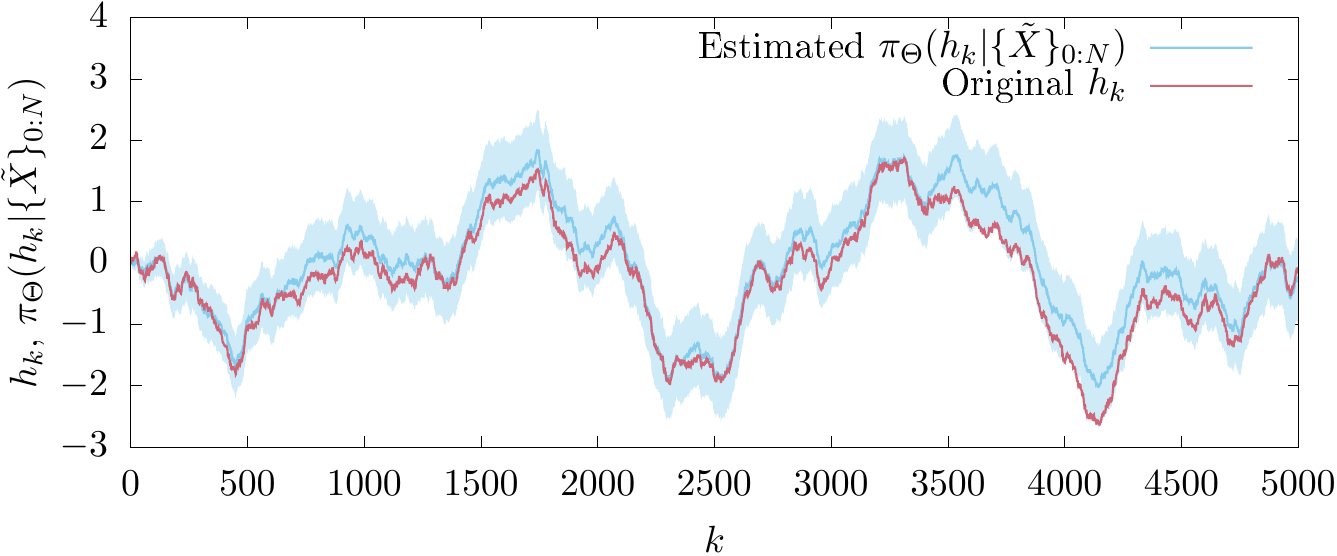}
  \caption{
  Estimation by the E-Step of the law of the trajectory of a single hidden variable ($d_h=1$), $\pi_{\Theta}({h_k}_{0:N}|\trajxv{0}{N} )$, reconstructed using only the trajectory of the observed variable $\trajxv{0}{N}$ for the one-dimensional case ($d=1$) with a quadratic potential $V(x)=x^2/2$ discussed in the main text (with more hidden variables).
  The figure shows the original trajectory of the hidden variable (red line) and the conditional law, represented by its mean $k\mapsto \avg{h_k}$ (blue line) and twice its standard deviation (shaded area).
  }
  \label{fig:Estep}
\end{figure}

\subsection*{E-step: explicit expressions}
Since the computations of the E-step follow from integrals over hidden variables, we decompose the average term in~\eqref{eq:euler_random_part} between a part that depends on $\tilde{X}_k$ and another that depends on the hidden variables,
\begin{equation*}
     \mu_{\Delta t}(\statec{k}) = \tilde{ \mu}(\tilde{X}_{k}) + {A_h} h_k \;,
\end{equation*} 
with ${A_h} $ a $ (d+d_h) \times d_h$ matrix.
The prediction-correction part of the E-step proceeds forward (iterating from $k-1$ to $k$) and we have for~\eqref{eq:forwardterm2}
\begin{equation}
\label{eq:filter_res}
     \prob{h_k| \trajxv{0}{k}} =  \mathcal{N}\left(h_k ;\mu^f_k,\Sigma^f_k\right)
\end{equation}
where the mean and variance are given by
\begin{eqnarray*}
\mu^f_k&=& \mu^*_h +\Sigma^*_{h,v}\left[\Sigma^*_{v,v}\right]^{-1}\left[v_k-\mu^*_v\right]   \\
\Sigma^f_k &=&  \Sigma^*_{h,h}-\Sigma^*_{h,v}\left[\Sigma^*_{v,v}\right]^{-1}\Sigma^*_{v,h}
\end{eqnarray*}
where
\begin{eqnarray*}
\begin{pmatrix} \mu^*_v \\ \mu^*_h\end{pmatrix} &= &\tilde{ \mu}(\tilde{X}_{k-1}) + {A_h} \mu^f_{k-1}  \\
 \begin{pmatrix} \Sigma^*_{v,v} & \Sigma^*_{v,h} \\\Sigma^*_{h,v} & \Sigma^*_{h,h}\end{pmatrix} &= & \sigma_{\Delta t}\sigma_{\Delta t}^\T + {A_h} \Sigma^f_{k-1} {A_h}^\T .
\end{eqnarray*}
The smoother part of the E-step proceeds backward (iterating from $k+1$ to $k$). Introducing the $  d_h \times (d+d_h)$ matrix
\begin{equation*}
    R = \begin{pmatrix} R_v & R_h\end{pmatrix}  = \Sigma^f_k {A_h}^\T \left[\sigma_{\Delta t}\sigma_{\Delta t}^\T + {A_h} \Sigma^f_k {A_h}^\T \right]^{-1} \;, 
\end{equation*}
we have for~\eqref{eq:pair_estimate_smoother}
\begin{equation}
\label{eq:smoother_res}
    \prob{h_k,h_{k+1}  | \trajxv{0}{N}}  = \mathcal{N}\left( \begin{pmatrix} h_k\\ h_{k+1}\end{pmatrix}    ; \begin{pmatrix} \mu^s_k \\ \mu^s_{k+1}\end{pmatrix}  , \begin{pmatrix}  \Sigma^s_k & R_h\Sigma^s_{k+1} \\ (R_h\Sigma^s_{k+1})^\T &  \Sigma^s_{k+1}\end{pmatrix} \right)
\end{equation}
using the expression of the marginal distribution $ \prob{h_{k} | \trajxv{0}{N}}= \mathcal{N}\left(  \mu^s_{k};\Sigma^s_{k} \right) $ whose mean and variance are
\begin{align*}
   \mu^s_k  & = \mu^f_k-R \left[{A_h}\mu^f_k+\tilde{ \mu}(\tilde{X}_{k})\right] +R_v v_{k+1}+R_h \mu^s_{k+1} \\
    \Sigma^s_k & =  R_h \Sigma^s_{k+1} R_h^\T + \Sigma^f_{k} - R{A_h}\Sigma^f_{k} .
\end{align*}

\subsection*{Comparison between the EM and Volterra methods and estimate of the (potential of) mean force}
In the main text we compare the EM and Volterra methods to compute the memory kernel from an initial set of trajectories. This requires an estimate of the potential of mean force (PMF) as a function of the collective variables. In the first two examples, we consider harmonic potentials in one and two dimensions, respectively. Figure~\ref{fig:comparison} compares the memory kernels obtained in the 1D case by the Volterra (V) and EM methods with various ways of computing the effective force. For both V and EM, we consider the kernels resulting from trajectories generated using the original harmonic potential used to generate the initial trajectories (original), as well as using the harmonic potential obtained by fitting the forces (for EM) or the potential of mean force (for V) sampled from the initial trajectories (fitted). In the V case, we also show the results for the default use of the memtools package which does not rely on a fit of the potential of mean force (obtained by histograms) by a harmonic potential but rather a numerical approximation by cubic splines. We find that there is little difference at long times between the original and fitted harmonic potentials, while using splines for the PMF with the Volterra method deteriorates the results compared to the fits by a quadratic potential (for V) or corresponding linear force (for EM). Nevertheless, even in these cases we observe some instability and a large variance at long times and the conclusions of the comparison between the proposed likelihood-based method and the Volterra ones are unchanged.

\begin{figure}[ht]
  \centering
  
  \includegraphics[width=0.7\linewidth]{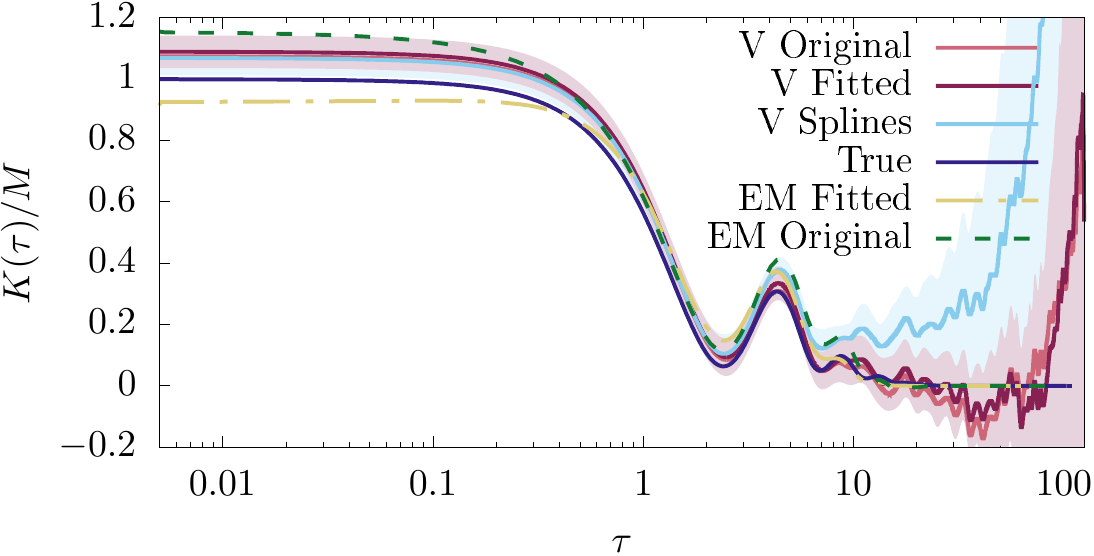}
  \caption{Comparison between memory kernels involved in the 1D example (see Fig.~2a of the main text). See the text of this Appendix for the description of the labels.
  }
  \label{fig:comparison}
\end{figure}

\end{widetext}

\bibliography{GLE}

\end{document}